# Achieving Practical OAM Based Wireless Communications With Misaligned Transceiver

Wenchi Cheng[†], Haiyue Jing[†], Wei Zhang[‡], Zan Li[†], and Hailin Zhang[†]

[†]State Key Laboratory of Integrated Services Networks, Xidian University, Xi'an, China
[‡]School of Electrical Engineering and Telecommunications, the University of New South Wales, Sydney, Australia
E-mail: {*wccheng@xidian.edu.cn, wzhang@ee.unsw.edu.au, zanli@xidian.edu.cn*, and *hlzhang@xidian.edu.cn*}

*Abstract*—Orbital angular momentum (OAM) has attracted much attention for radio vortex wireless communications due to the orthogonality among different OAM-modes. To maintain the orthogonality among different OAM modes at the receiver, the strict alignment between transmit and receive antennas is highly demanded. However, it is not practical to guarantee the transceiver alignment in wireless communications. The phase turbulence, resulting from the misaligned transceivers, leads to serious inter-mode interference among different OAM modes and therefore fail for signals detection of multiple OAM modes at the receiver. To achieve practical OAM based wireless communications, in this paper we investigate the radio vortex wireless communications with misaligned transmit and receive antennas. We propose a joint Beamforming and Pre-detection (BePre) scheme, which uses two unitary matrices to convert the channel matrix into the equivalent circulant matrix for keeping the orthogonality among OAM-modes at the receiver. Then, the OAM signals can be detected with the mode-decomposition scheme at the misaligned receiver. Extensive simulations obtained validate and evaluate that our developed joint BePre scheme can efficiently detect the signals of multiple OAM-modes for the misaligned transceiver and can significantly increase the spectrum efficiency.

## I. INTRODUCTION

DURING the past several years, orbital angular momentum (OAM), which is an important property of electromagnetic wave, has received much attention [1]–[4]. The electromagnetic wave carrying OAM contains an infinite number of topological charges, i.e., the OAM-modes. Different OAM-modes are orthogonal with each other. It is expected to use the orthogonality among different OAM-modes for wireless communications.

OAM signals can be generated and detected by helical phase plate antenna [5], helical paraboloid antenna [6], and uniform circular array (UCA) antenna [7], etc. Due to its flexibility for digitally generating multiple OAM-modes simultaneously, the UCA antenna is extensively studied and highly expected to be used for OAM based wireless communications. The authors of [8] showed that UCA is superior to radial array and tangential array from the perspective of multiplexing and demultiplexing. The authors of [9] analyzed the UCA configuration for generating OAM beams and analyzed the impact of array error on the radiation field. The transmission characteristics of multiple OAM-modes generated by UCA antenna for different distances are investigated [10]. The authors of [11] investigated the UCA antenna based OAM transmission for wireless communications in the aspect of orthogonality, degree of freedom, and capacity. For applications, mode jamming [12] and index modulation [13] for OAM based wireless communications are also investigated.

Most existing schemes are designed for the transceiver-aligned scenario where different OAM-modes can be easily distinguished at the receiver [11], [12]. However, it is difficult to ensure that transmit and receive UCAs are aligned with each other in practical wireless communications scenarios [14]. If the transmit and receive UCAs are not aligned, it is difficult to decompose the signals with different OAM-modes due to the phase turbulence resulting from the unequal distance transmission [4]. The authors of [15] showed that the bit error rate of OAM system is highly dependent on the alignment between transmit and receive UCAs. The authors of [16] investigated the impact of misalignment on the channel capacity and proposed a beam steering scheme to avoid the sharp performance degradation. However, the capacity is still significantly decreased as compared with that of the aligned scenario [16]. Thus, a question is raised that how to decompose the OAM beams with multiple OAM-modes for the misaligned transmit and receive UCAs scenario while still achieving high spectrum efficiency.

To overcome the above-mentioned problem, in this paper we propose a joint Beamforming and Pre-detection (BePre) scheme for the OAM based radio vortex wireless communications with misaligned transmit and receive UCAs. The beamforming and pre-detection matrices are unitary and can convert the channel matrix into the equivalent circulant matrix. What we need is just the channel state information at the transmitter and receiver. Then, the OAM signals can be detected with low complexity based maximum likelihood (ML) detection at the misaligned receiver. The obtained numerical results show that our proposed scheme can decompose the signals carried by multiple OAM-modes with low complexity and can significantly increase the spectrum efficiency for OAM based wireless communications using the joint BePre scheme as compared with that for OAM based wireless communications without using the joint BePre scheme.

This work was supported in part by the National Natural Science Foundation of China (No. 61771368), the Young Elite Scientists Sponsorship Program by CAST (2016QNRC001), the Young Talent Support Fund of Science and Technology of Shaanxi Province (2018KJXX-025), and the Australian Research Council's Projects funding scheme under Projects (DP160104903).

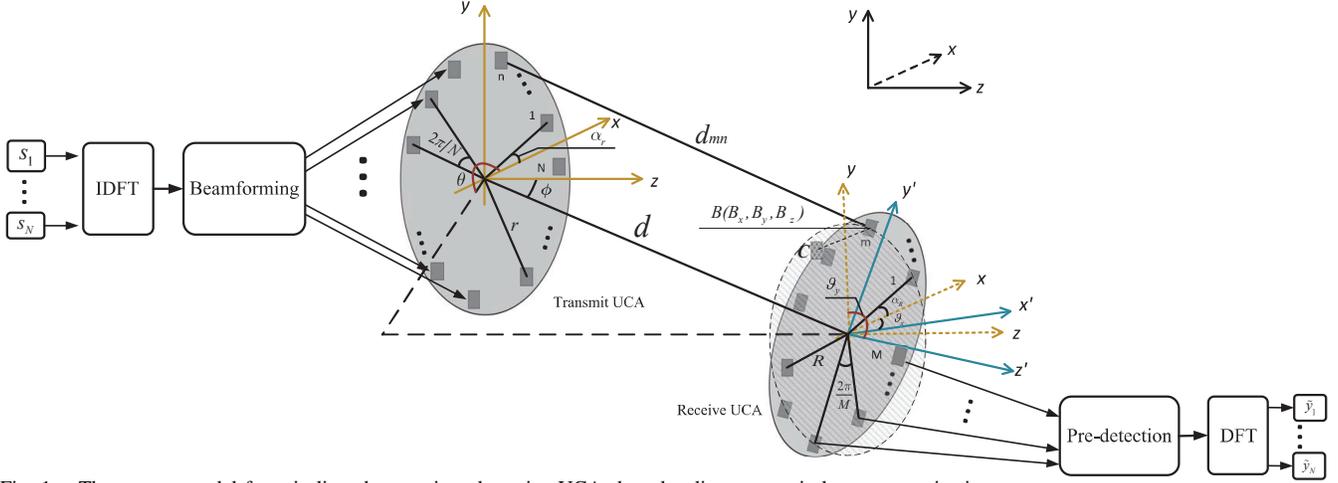

Fig. 1. The system model for misaligned transmit and receive UCAs based radio vortex wireless communications.

The rest of this paper is organized as follows. Section II gives the misaligned transmit and receive UCAs based radio vortex wireless communications model. Section III develops the joint BePre scheme and the mode-decomposition scheme as well as derives the spectrum efficiency for the misaligned transmit and receive UCAs scenario. Section IV evaluates our proposed schemes and the obtained spectrum efficiency versus the included angles as well as the rotation angles. The paper concludes with Section V.

## II. THE MISALIGNED UCAS BASED SYSTEM MODEL FOR RADIO VORTEX WIRELESS COMMUNICATION

Figure 1 depicts the system model for the misaligned transmit and receive UCAs based radio vortex wireless communications. We denote by $r$ and $R$ the radius of transmit UCA and the radius of receive UCA, respectively. The sizes of transmit and receive UCAs can be different. We denote by $d$ the distance from the center of the transmit UCA to the center of the receive UCA. The transmit and receive UCAs are equipped with $N$ antenna elements and $M$ antenna elements, respectively. For the transmit UCA, the antenna elements, uniformly around the perimeter of the circle, are fed with the same input signal but with a successive delay from antenna element to antenna element such that after a full turn the phase has been incremented by $2\pi l$, where $l$ represents the number of topological charges, i.e., OAM-modes. For the receive UCA, the antenna elements are also uniformly around perimeter of the circle. We denote by $\alpha_r$ and $\alpha_R$ the angles between the phase angle of the first antenna element and zero radian corresponding to the transmit and receive UCAs, respectively. The parameter $\theta$ denotes the included angle between $x$-axis and the projection of the line from the center of the transmit UCA to the center of the receive UCA on the transmit plane. Also, $\phi$ denotes the included angle between $z$-axis and the line from the center of the transmit UCA to the center of the receive UCA. The rotation angle $\vartheta_y$ represents the angle of rotation around the $x$-axis. The rotation angle $\vartheta_x$ represents the angle of rotation around the $y$-axis.

At the transmitter, the transmit signals become vorticose using the inverse discrete Fourier transform (IDFT). After the beamforming, the signals are emitted by transmit UCA. The received signals first pre-detect at the receiver and then, the transmit signals can be recovered using the discrete Fourier transform (DFT). In the following, we propose the joint BePre scheme for misaligned transmit and receive UCAs based radio vortex wireless communications. We then decompose the signals on each mode and analyze the achieved spectrum efficiency.

## III. JOINT BEPRE SCHEME FOR MISALIGNED UCAS BASED RADIO VORTEX WIRELESS COMMUNICATIONS

The transmit signal of the $n$th $(1 \leq n \leq N)$ antenna element on the transmit UCA, denoted by $x_n$, is given as follows:

$$\begin{aligned} x_n &= \sum_{l=\lfloor \frac{2-N}{2} \rfloor}^{\lfloor N/2 \rfloor} \frac{1}{\sqrt{N}} s_l e^{j(\varphi_n + \alpha_r)l} \\ &= \sum_{l=\lfloor \frac{2-N}{2} \rfloor}^{\lfloor N/2 \rfloor} \frac{1}{\sqrt{N}} s_l e^{j\left[\frac{2\pi(n-1)}{N} + \alpha_r\right]l}, \end{aligned} \quad (1)$$

where $\varphi_n = 2\pi(n-1)/N$ is the basic rotation-angle for the transmit UCA and $\varphi_n + \alpha_r$ is the azimuthal angle, defined as the angular position on a plane perpendicular to the axis of propagation, for the $n$th antenna element on the transmit UCA. The symbol $s_l$ denotes the transmit signal corresponding to the $l$th OAM-mode of the transmit UCA and the symbol $\lfloor \kappa \rfloor$ represents the largest integer which is less than the integer $\kappa$.

We denote by $h_{mn}$ the channel gain from the $n$th antenna element on the transmit UCA to the $m$th $(1 \leq m \leq M)$ antenna element on the receive UCA. Then, $h_{mn}$ can be written as follows [17]:

$$h_{mn} = \frac{\beta \lambda e^{-j\frac{2\pi}{\lambda} d_{mn}}}{4\pi d_{mn}}, \quad (2)$$

where $\beta$ denotes the combination of all the relevant constants such as attenuation and phase rotation caused by antennas and their patterns on both sides [17]. The parameter $d_{mn}$ represents the distance between the $n$th antenna element on the transmit UCA and the $m$th antenna element on the receive UCA.

$$\begin{cases}
\mathcal{E}_m = dR\sin\phi\cos\theta\cos(\psi_m + a_R)\cos\vartheta_y + dR\sin\phi\cos\theta\sin(\psi_m + a_R)\sin\vartheta_x\sin\vartheta_y \\
\quad + dR\sin\phi\sin\theta\sin(\psi_m + a_R)\cos\vartheta_x + dR\cos\phi\cos(\psi_m + a_R)\sin\vartheta_y - dR\cos\phi\sin(\psi_m + a_R)\sin\vartheta_x\cos\vartheta_y; \\
\mathcal{D}_{mn} = -Rr\cos(\psi_m + a_R)\cos(\varphi_n + \alpha_r)\cos\vartheta_y - Rr\sin(\psi_m + a_R)\cos(\varphi_n + \alpha_r)\sin\vartheta_x\sin\vartheta_y \\
\qquad\qquad\qquad\qquad\qquad\qquad\qquad\qquad\qquad\qquad\qquad -Rr\sin(\psi_m + a_R)\sin(\varphi_n + \alpha_r)\cos\vartheta_x; \\
\mathcal{F}_n = -dr\sin\phi\cos(\varphi_n + \alpha_r - \theta)
\end{cases} \quad (5)$$

The coordinate of the $n$th antenna element on the transmit UCA is $(r\cos(\varphi_n + \alpha_r), r\sin(\varphi_n + \alpha_r), 0)$ and the coordinate corresponding to the center of the receive UCA is $(d\sin\phi\cos\theta, d\sin\phi\sin\theta, d\cos\phi)$. Similar to the azimuthal angle at the transmit UCA, $\psi_m = 2\pi(m-1)/M$ is the basic rotation-angle for the receive UCA and $\psi_m + a_R$ is the azimuthal angle of the $m$th antenna element on the receive UCA.

*Lemma 1:* The coordinate of the $m$th antenna element, denoted by $(B_x, B_y, B_z)$, on the receive UCA is given as follows:

$$\begin{cases}
B_x = d\sin\phi\cos\theta + R\cos(\psi_m + a_R)\cos\vartheta_y \\
\qquad + R\sin(\psi_m + a_R)\sin\vartheta_x\sin\vartheta_y; \\
B_y = d\sin\phi\sin\theta + R\sin(\psi_m + a_R)\cos\vartheta_x; \\
B_z = d\cos\phi + R\cos(\psi_m + a_R)\sin\vartheta_y \\
\qquad - R\sin(\psi_m + a_R)\sin\vartheta_x\cos\vartheta_y.
\end{cases} \quad (3)$$

*Proof:* See Appendix A. ∎

Then, the distance between the $n$th antenna element on the transmit UCA and the $m$th antenna element on the receive UCA, denoted by $d_{mn}$, can be derived as follows:

$$d_{mn} = \sqrt{[B_x - r\cos(\varphi_n + \alpha_r)]^2 + [B_y - r\sin(\varphi_n + \alpha_r)]^2 + B_z^2}$$
$$= \sqrt{d^2 + R^2 + r^2 + 2\mathcal{E}_m + 2\mathcal{D}_{mn} + 2\mathcal{F}_n}, \quad (4)$$

where $\mathcal{E}_m$, $\mathcal{D}_{mn}$ and $\mathcal{F}_n$ are given in Eq. (5).

Taking Eq. (4) into Eq. (2), we can obtain the specified channel gain and the channel matrix, denoted by $\boldsymbol{H}$. Clearly, the channel matrix $\boldsymbol{H}$ is not a circulant matrix. We denote by $\boldsymbol{s} = [s_{\lfloor\frac{2-N}{2}\rfloor}, \cdots, s_{\lfloor\frac{N}{2}\rfloor}]^T$ the basic transmit signal vector at the transmit UCA. Then, the transmit vortex signal vector, denoted by $\boldsymbol{x} = [x_1, \cdots, x_N]^T$, is given as follows:

$$\boldsymbol{x} = \boldsymbol{W}\boldsymbol{s}, \quad (6)$$

where the matrix $\boldsymbol{W} = \left\{\frac{1}{\sqrt{N}}\exp[2\pi(n-1)(l-1)/N]\right\}$ is the standard $N \times N$ IDFT matrix with $1 \leq n \leq N$ and $\lfloor\frac{2-N}{2}\rfloor \leq l \leq \lfloor\frac{N}{2}\rfloor$. Then, we have the received signal vector, denoted by $\boldsymbol{y} = [y_1, \cdots, y_N]^T$, as follows:

$$\boldsymbol{y} = \boldsymbol{H}\boldsymbol{W}\boldsymbol{s} + \boldsymbol{z}, \quad (7)$$

where $\boldsymbol{z} = [z_1, \cdots, z_N]^T$ is the noise vector at the receiver.

It is very difficult to directly decompose the transmit signals using DFT matrix because the channel matrix $\boldsymbol{H}$ is not a circulant matrix. For conveniently using DFT matrix at the receiver, we propose the joint BePre scheme to convert the channel matrix into the equivalent circulant matrix, for which we assume that the number of antenna elements on the transmit UCA is equal to that on the receive UCA, i.e., $N = M$.

### A. The Joint BePre Scheme

When the transmit and receive UCAs are aligned with each other, the channel matrix $\boldsymbol{H}$ is a circulant matrix. Then, after the DFT, the computation complexity of recovering the transmit signals at the receive UCA is very low since the ML based multiple OAM modes detection can be transformed into independent singular OAM mode detection. However, for practical wireless communications with generic scenarios, the channel matrix is not a circulant matrix becasue the transmit and the receive UCAs are not aligned. Thus, it is highly demanded to convert the channel matrix into a equivalent circulant matrix using unitary beamforming and pro-detection matrices when the transmit and receive UCAs are misaligned. In the following, we give the steps on how to derive the circulant matrix, the beamforming matrix, and the pre-detection matrix. What we need is just the channel estimation results.

**Step 1:** Derive the circulant matrix.

We denoted by $\boldsymbol{H}^c$ the circualnt matrix. Using the singular value decomposition for $\boldsymbol{H}$ and $\boldsymbol{H}^c$, we have $\boldsymbol{H} = \boldsymbol{S_H}\boldsymbol{V_H}\boldsymbol{U_H^*}$ and $\boldsymbol{H}^c = \boldsymbol{S_{H^c}}\boldsymbol{V_{H^c}}\boldsymbol{U_{H^c}^*}$, where the matrices $\boldsymbol{S_H}$, $\boldsymbol{U_H}$, $\boldsymbol{S_{H^c}}$, and $\boldsymbol{U_{H^c}}$ are unitary matrices. The notation $(\cdot)^*$ represents the conjugate transpose operation. The matrices $\boldsymbol{V_H}$ and $\boldsymbol{V_{H^c}}$, which are diagonal matrices, are singular value matrices corresponding to $\boldsymbol{H}$ and $\boldsymbol{H}^c$, respectively. To avoid signal-to-noise ratio (SNR) reduction, the singular value matrix of OAM based wireless communications using the joint BePre scheme should be equal to that of OAM based wireless communications without using the joint BePre scheme. Thus, let $\boldsymbol{V_H}$ be equal to $\boldsymbol{V_{H^c}}$.

Since the sequence of diag($\boldsymbol{V_{H^c}}$), denoted by $[h_1^c, h_2^c, \cdots, h_N^c]$, is the N-point DFT of the first row corresponding to the circulant matrix $\boldsymbol{H}^c$, we have

$$\sqrt{N}\boldsymbol{W}^*[h_1^c, h_2^c, \cdots, h_N^c]^T = \text{diag}(\boldsymbol{V_{H^c}}). \quad (8)$$

Also, because the matrix $\boldsymbol{W}^*$ is a full rank matrix, the sequence $[h_1^c, h_2^c, \cdots, h_N^c]$ can be derived as follows:

$$[h_1^c, h_2^c, \cdots, h_N^c] = \left[\frac{1}{\sqrt{N}}\boldsymbol{W}\,\text{diag}(\boldsymbol{V_{H^c}})\right]^T. \quad (9)$$

Based on the first row of the circulant matrix, we can obtain the circulant matrix $\boldsymbol{H}^c$. Then, the unitary matrices $\boldsymbol{S_{H^c}}$ and $\boldsymbol{U_{H^c}}$ can be derived using the singular value decomposition.

**Step 2:** Derive the beamforming matrix and the pre-detection matrix.

We denoted by $\boldsymbol{H}^{Pt}$ and $\boldsymbol{H}^{Pr}$ the beamforming matrix and the pre-detection matrix, respectively. To obtain the beamforming matrix $\boldsymbol{H}^{Pt}$ and the pre-detection matrix $\boldsymbol{H}^{Pr}$, we use $\boldsymbol{H}^{Pt} = \boldsymbol{S}_{\boldsymbol{H}^{Pt}} \boldsymbol{V}_{\boldsymbol{H}^{Pt}} \boldsymbol{U}^*_{\boldsymbol{H}^{Pt}}$ and $\boldsymbol{H}^{Pr} = \boldsymbol{S}_{\boldsymbol{H}^{Pr}} \boldsymbol{V}_{\boldsymbol{H}^{Pr}} \boldsymbol{U}^*_{\boldsymbol{H}^{Pr}}$, which are the commonly used singular value decomposition. Since $\boldsymbol{H}^c = \boldsymbol{H}^{Pr} \boldsymbol{H} \boldsymbol{H}^{Pt}$, we have

$$\boldsymbol{S}_{\boldsymbol{H}^c} \boldsymbol{V}_{\boldsymbol{H}^c} \boldsymbol{U}^*_{\boldsymbol{H}^c}
= \boldsymbol{S}_{\boldsymbol{H}^{Pr}} \boldsymbol{V}_{\boldsymbol{H}^{Pr}} \boldsymbol{U}^*_{\boldsymbol{H}^{Pr}} \boldsymbol{S}_{\boldsymbol{H}} \boldsymbol{V}_{\boldsymbol{H}} \boldsymbol{U}^*_{\boldsymbol{H}} \boldsymbol{S}_{\boldsymbol{H}^{Pt}} \boldsymbol{V}_{\boldsymbol{H}^{Pt}} \boldsymbol{U}^*_{\boldsymbol{H}^{Pt}}. \quad (10)$$

Specially, we have $\boldsymbol{S}_{\boldsymbol{H}^c} = \boldsymbol{S}_{\boldsymbol{H}^{Pr}}$, $\boldsymbol{U}_{\boldsymbol{H}^{Pr}} = \boldsymbol{S}_{\boldsymbol{H}}$, $\boldsymbol{U}_{\boldsymbol{H}} = \boldsymbol{S}_{\boldsymbol{H}^{Pt}}$, and $\boldsymbol{U}_{\boldsymbol{H}^c} = \boldsymbol{U}_{\boldsymbol{H}^{Pt}}$. Then, we can obtain

$$\boldsymbol{V}_{\boldsymbol{H}^c} = \boldsymbol{V}_{\boldsymbol{H}^{Pr}} \boldsymbol{V}_{\boldsymbol{H}} \boldsymbol{V}_{\boldsymbol{H}^{Pt}}. \quad (11)$$

If the beamforming matrix or the pre-detection matrix is not an unitary matrix, the power corresponding to signals with different OAM-modes will change, resulting in small spectrum efficiency. Thus, both the beamforming matrix and the pre-detection matrix should be unitary matrices. Since $\boldsymbol{S}_{\boldsymbol{H}^{Pr}}$, $\boldsymbol{U}_{\boldsymbol{H}^{Pr}}$, $\boldsymbol{S}_{\boldsymbol{H}^{Pt}}$, and $\boldsymbol{U}_{\boldsymbol{H}^{Pr}}$ are unitary matrices, the diagonal matrices $\boldsymbol{V}_{\boldsymbol{H}^{Pr}}$ and $\boldsymbol{V}_{\boldsymbol{H}^{Pt}}$ are unitary matrices as well. Specially, both the diagonal matrix $\boldsymbol{V}_{\boldsymbol{H}^{Pr}}$ and the diagonal matrix $\boldsymbol{V}_{\boldsymbol{H}^{Pt}}$ are identity matrices. Then, the beamforming matrix $\boldsymbol{H}^{Pt}$ and the pre-detection matrix $\boldsymbol{H}^{Pr}$ can be obtained as follows:

$$\begin{cases} \boldsymbol{H}^{Pt} = \boldsymbol{U}_{\boldsymbol{H}} \boldsymbol{U}^*_{\boldsymbol{H}^c}; \\ \boldsymbol{H}^{Pr} = \boldsymbol{S}_{\boldsymbol{H}^c} \boldsymbol{S}^*_{\boldsymbol{H}}. \end{cases} \quad (12)$$

### B. Mode-Decomposition and Spectrum Efficiency Analyses

After the beamforming, we have the transmit signal vector as $\widetilde{\boldsymbol{x}} = \boldsymbol{H}^{Pt} \boldsymbol{W} \boldsymbol{s}$. The received signal, denoted by $\boldsymbol{y}$, can be given as follows:

$$\boldsymbol{y} = \boldsymbol{H} \widetilde{\boldsymbol{x}} + \boldsymbol{z} = \boldsymbol{H} \boldsymbol{H}^{Pt} \boldsymbol{W} \boldsymbol{s} + \boldsymbol{z}. \quad (13)$$

Multiplying $\boldsymbol{W}^* \boldsymbol{H}^{Pr}$ with the received signal vector $\boldsymbol{y}$, we have

$$\begin{aligned}
\boldsymbol{W}^* \boldsymbol{H}^{Pr} \boldsymbol{y} &= \boldsymbol{W}^* \boldsymbol{H}^{Pr} \boldsymbol{H} \boldsymbol{H}^{Pt} \boldsymbol{W} \boldsymbol{s} + \boldsymbol{W}^* \boldsymbol{H}^{Pr} \boldsymbol{z} \\
&= \boldsymbol{W}^* \boldsymbol{H}^c \boldsymbol{W} \boldsymbol{s} + \boldsymbol{W}^* \boldsymbol{H}^{Pr} \boldsymbol{z} \\
&= \boldsymbol{\Lambda} \boldsymbol{s} + \boldsymbol{W}^* \boldsymbol{H}^{Pr} \boldsymbol{z}, \quad (14)
\end{aligned}$$

where $\boldsymbol{\Lambda} = \text{diag}(\Lambda_1, \Lambda_2, \cdots, \Lambda_N)$ is a diagonal matrix corresponding to the circulant matrix $\boldsymbol{H}^c$. Denoting by $\widetilde{\boldsymbol{y}} = \boldsymbol{W}^* \boldsymbol{H}^{Pr} \boldsymbol{y} = [\widetilde{y}_1, \widetilde{y}_2, \cdots, \widetilde{y}_N]^T$ and using ML detection scheme, we can obtain the estimate signal, denoted by $\widehat{\boldsymbol{s}}$, as follows:

$$\begin{aligned}
\widehat{\boldsymbol{s}} &= \arg \min_{\boldsymbol{s} \in \boldsymbol{\Omega}^N} \|\widetilde{\boldsymbol{y}} - \boldsymbol{\Lambda} \boldsymbol{s}\|^2 \\
&= \arg \min_{\boldsymbol{s} \in \boldsymbol{\Omega}^N} \sum_{i=1}^N |\widetilde{y}_i - \Lambda_i s_i|^2 \\
&= \left[ \arg \min_{s_1 \in \boldsymbol{\Omega}} |\widetilde{y}_1 - \Lambda_1 s_1|^2, \cdots, \arg \min_{s_N \in \boldsymbol{\Omega}} |\widetilde{y}_N - \Lambda_N s_N|^2 \right]^T \\
&= \left[ \arg \min_{s_1 \in \boldsymbol{\Omega}} |\widetilde{y}_1 - \Lambda_1 s_1|, \cdots, \arg \min_{s_N \in \boldsymbol{\Omega}} |\widetilde{y}_N - \Lambda_N s_N| \right]^T, \quad (15)
\end{aligned}$$

where $\boldsymbol{\Omega}$ is the signal constellation with size $\xi$ (the number of modulation constellation in $\boldsymbol{\Omega}$).

For OAM based wireless communications without using the joint BePre scheme, if the channel matrix is not a circulant matrix, there exists interference among signals with different OAM-modes, i.e., the OAM-modes are not orthogonal with each other. The spectrum efficiency, denoted by $C_T$, of OAM based wireless communications without using the joint BePre scheme can be derived as follows:

$$C_T = \sum_{i=1}^N \log_2 \left( 1 + \frac{|\widetilde{h}_{ii}|^2 P_i}{\sigma_i^2 + \sum_{k=1, k \neq i}^N |\widetilde{h}_{ik}|^2 P_k} \right), \quad (16)$$

where $\widetilde{h}_{ik}$ represents the element of $i$th ($1 \leq i \leq N$) row and $k$th ($1 \leq k \leq N$) column corresponding to the matrix $\boldsymbol{W}^* \boldsymbol{H} \boldsymbol{W}$, $P_i$ denotes the power allocated to the $i$th signal, and $\sigma_i^2$ denotes the variance of received noise.

For OAM based wireless communications using the joint BePre scheme, the channel matrix is converted into the equivalent circulant matrix and the signals with different OAM-modes can be considered interference-free. We can derive the spectrum efficiency, denoted by $C_c$, of OAM based wireless communications using the joint BePre scheme as follows:

$$C_c = \sum_{i=1}^{rank(\boldsymbol{H})} \log_2 \left( 1 + \frac{\gamma_i^{\boldsymbol{H}^c} P_i}{\widetilde{\sigma}_i^2} \right) = \sum_{i=1}^{rank(\boldsymbol{H})} \log_2 \left( 1 + \frac{\gamma_i^{\boldsymbol{H}} P_i}{\widetilde{\sigma}_i^2} \right), \quad (17)$$

where $\gamma_i^{\boldsymbol{H}^c}$ represents the singular values corresponding to the circulant matrix and $\widetilde{\sigma}_i^2$ is given by

$$\widetilde{\sigma}_i^2 = \sum_{k=1}^N \left| h_{ik}^{\boldsymbol{H}^{Pr}} \right|^2 \sigma_k^2. \quad (18)$$

In Eq. (18), $h_{ik}^{\boldsymbol{H}^{Pr}}$ denotes the elements of the pre-detection matrix $\boldsymbol{H}^{Pr}$. Since the pre-detection matrix $\boldsymbol{H}^{Pr}$ is an unitary matrix, the variance of the received noise remains unchange.

## IV. Performance Evaluations

In this section, we numerically evaluate OAM based wireless communications using the joint BePre scheme for the practical scenario where the transmit and receive UCAs are misaligned. We also compare the number of computations and maximum spectrum efficiency corresponding to OAM based wireless communications using the joint BePre scheme with those of OAM based wireless communications without using the joint BePre scheme. The maximum spectrum efficiency is obtained by using water-filling power allocation scheme. Throughout our evaluations, we set the wavelength $\lambda$ as 0.01 m. We also set $R = r = 4\lambda$, $\beta = 1$, and $\alpha_r = \alpha_R = 0$.

Figure 2 shows the numbers of additions and multiplications for OAM based wireless communications using the joint BePre scheme and OAM based wireless communications without using the joint BePre scheme, where we set $\xi = 4$. The numbers of additions and multiplications increase as the number of antenna elements increases. This is because $\xi$ increases as the number of antenna elements increases. Also, we can observe that the numbers of additions and multiplications for OAM based wireless communications without using the joint

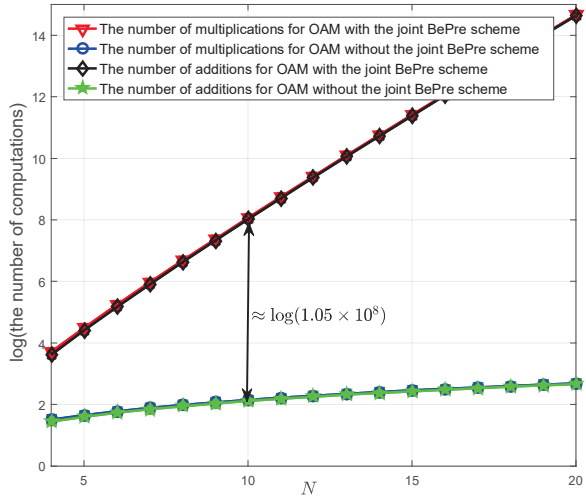

Fig. 2. The numbers of computations for OAM based wireless communications with the joint BePre scheme and OAM based wireless communications without the joint BePre scheme.

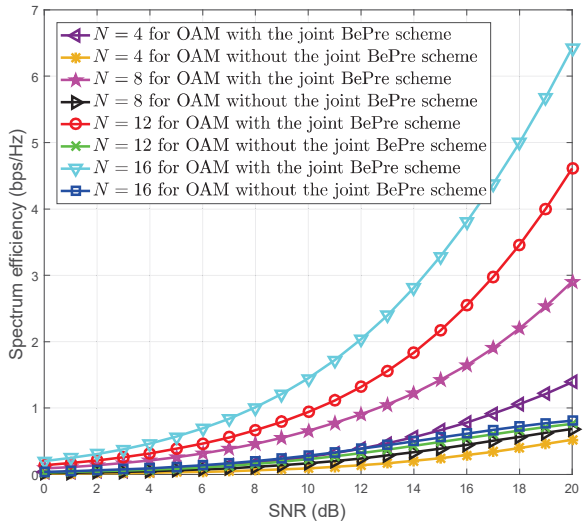

Fig. 3. The spectrum efficiencies for OAM based wireless communications with the joint BePre scheme and OAM based wireless communications without the joint BePre scheme.

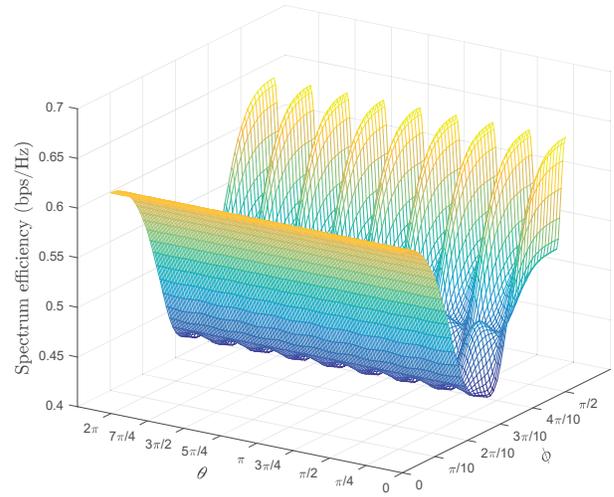

Fig. 4. The obtained spectrum efficiency versus the included angles $\phi$ and $\theta$.

BePre scheme are much larger than those of OAM based wireless communications using the joint BePre scheme. For example, the gap between the number of additions for OAM based wireless communications without using the joint BePre scheme and the number of additions for OAM based wireless communications using the joint BePre scheme is approximately $1.05 \times 10^8$ when $N = 10$. This is because the ML based multiple OAM-modes detection can be transformed into independent singular OAM-mode detection, thus significantly reducing the computational complexity.

Figure 3 depicts the spectrum efficiencies of OAM based wireless communications using the joint BePre scheme and OAM based wireless communications without using the joint BePre scheme, where the transmit and receive UCA are misaligned. We can obtain that the spectrum efficiency of OAM based wireless communications using the joint BePre scheme is much larger than that of OAM based wireless communications without using the joint BePre scheme. This is because the beamforming and pre-detection matrices both are unitary and they do not change the power of the signals. Using the beamforming and pre-detection matrices, the signals carried by different OAM-modes are interference-free. The spectrum efficiencies of OAM based wireless communications using the joint BePre scheme and OAM based wireless communications without using the joint BePre scheme increase as the number of antenna elements increases.

Figure 4 illustrates the spectrum efficiency versus the included angles $\phi$ and $\theta$, where we set $N = 8$, $\vartheta_x = 0$, and $\vartheta_y = 0$. when $\phi$ is smaller than $\pi/5$, the spectrum efficiencies corresponding to different $\theta$ for OAM based wireless communications using the joint BePre scheme are equal. This indicates that the included angle $\theta$ has no impact on the spectrum efficiencies. The spectrum efficiencies of OAM based wireless communications using the joint BePre scheme first decrease and then increase as the included angle $\phi$ increases.

Figure 5 displays the spectrum efficiencies of OAM based wireless communications using the joint BePre scheme and OAM based wireless communications without using the joint BePre scheme versus rotation angles $\vartheta_x$ and $\vartheta_y$, where we set $N = 8$, $\phi = 0$, and $\theta = 0$. We can observe that the spectrum efficiencies of OAM based wireless communications using the joint BePre scheme is very low when $\pi/5 < \vartheta_x, \vartheta_y < 3\pi/10$. The spectrum efficiency for OAM based wireless communications using the joint BePre scheme first decreases and then increases as the rotation angle $\vartheta_y$ increases when $\vartheta_x < 3\pi/10$.

## V. CONCLUSIONS

In this paper, we proposed the joint BePre scheme for the OAM based radio vortex wireless communications where the transmit and receive UCAs are misaligned. Using the joint BePre scheme, the channel matrix can be transformed into the equivalent circulant matrix, which has the same singular value matrix with the original channel matrix. Then, we developed the mode-decomposition scheme to recover the transmit signals as well as analyzed the obtained spectrum efficiency.

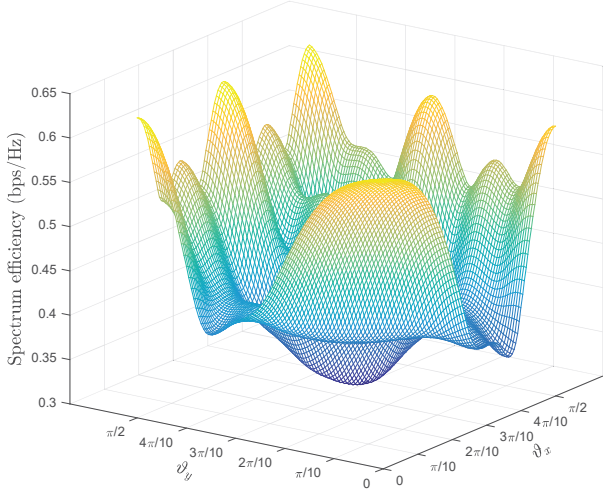

Fig. 5. The obtained spectrum efficiency versus the rotation angles $\vartheta_x$ and $\vartheta_y$.

Simulation results validated that the spectrum efficiency of OAM based wireless communications using the joint BePre scheme is much larger than that of OAM based wireless communications without using the joint BePre scheme for practical misaligned transceiver based radio vortex wireless communications and our proposed joint BePre scheme can significantly decrease the number of computations using ML detection.

## APPENDIX A
## PROOF OF LEMMA 1

For the generic scenario where the transmit and receive UCAs are misaligned, if the center of receive UCA is overlapped with the center of the coordinate system, we can obtain the coordinate of the $m$th antenna element, denoted by $(B'_x, B'_y, B'_z)$, as follows:

$$\begin{bmatrix} B'_x \\ B'_y \\ B'_z \end{bmatrix} = [\boldsymbol{R}_x(\vartheta_y)\boldsymbol{R}_y(\vartheta_x)]^T \begin{bmatrix} R\cos(\psi_m + \alpha_R) \\ R\sin(\psi_m + \alpha_R) \\ 0 \end{bmatrix} \quad (19)$$

where $\boldsymbol{R}_x(\vartheta_y)$ and $\boldsymbol{R}_y(\vartheta_x)$ represent the attitude matrices corresponding to the $x$-axis and $y$-axis, respectively. The attitude matrices $\boldsymbol{R}_x(\vartheta_y)$ and $\boldsymbol{R}_y(\vartheta_x)$ are given by

$$\boldsymbol{R}_x(\vartheta_y) = \begin{bmatrix} 1 & 0 & 0 \\ 0 & \cos\vartheta_y & \sin\vartheta_y \\ 0 & -\sin\vartheta_y & \cos\vartheta_y \end{bmatrix} \quad (20)$$

and

$$\boldsymbol{R}_y(\vartheta_x) = \begin{bmatrix} \cos\vartheta_x & 0 & -\sin\vartheta_x \\ 0 & 1 & 0 \\ \sin\vartheta_x & 0 & \cos\vartheta_x \end{bmatrix}, \quad (21)$$

respectively.

Thus, for the coordinate $(d\sin\phi\cos\theta, d\sin\phi\sin\theta, d\cos\phi)$, which is the center of the receive UCA, we can derive the coordinate corresponding to the $m$th antenna element on the receive UCA as follows:

$$\begin{bmatrix} B_x \\ B_y \\ B_z \end{bmatrix} = \begin{bmatrix} B'_x \\ B'_y \\ B'_z \end{bmatrix} + \begin{bmatrix} d\sin\phi\cos\theta \\ d\sin\phi\sin\theta \\ d\sin\phi \end{bmatrix}. \quad (22)$$

In summary, the coordinate of the $m$th antenna element on the receive UCA is shown in Eq. (3).